# Online Popularity under Promotion: Viral Potential, Forecasting, and the Economics of Time

**Marian-Andrei Rizoiu** and **Lexing Xie**
The Australian National University and Data61 CSIRO, Canberra, Australia.

**Abstract**

Modeling the popularity dynamics of an online item is an important open problem in computational social science. This paper presents an in-depth study of popularity dynamics under external promotions, especially in predicting popularity jumps of online videos, and determining effective and efficient schedules to promote online content. The recently proposed Hawkes Intensity Process (HIP) models popularity as a non-linear interplay between exogenous stimuli and the endogenous reactions. Here, we propose two novel metrics based on HIP: to describe popularity gain per unit of promotion, and to quantify the time it takes for such effects to unfold. We make increasingly accurate forecasts of future popularity by including information about the intrinsic properties of the video, promotions it receives, and the non-linear effects of popularity ranking. We illustrate by simulation the interplay between the unfolding of popularity over time, and the time-sensitive value of resources. Lastly, our model lends a novel explanation of the commonly adopted periodic and constant promotion strategy in advertising, as increasing the perceived viral potential. This study provides quantitative guidelines about setting promotion schedules considering content virality, timing, and economics.

## 1 Introduction

"The fundamental scarcity in the modern world is the scarcity of attention." – Herbert A. Simon. Human attention is limited, both for individuals and groups, and the mechanisms governing its allocation still remain largely not understood. Influencing attention allocation is a related open problem that has broad applications, such as optimizing information dissemination for producers and managing information overload for consumers. In this paper, we study the popularity of cultural items under promotion in the online environment. We seek to answer important questions on social attention, such as: how well will an item respond to a given amount of promotion? How much promotion is required for this item to rise to the top 5% in popularity? What is the best timing for spending a promotion budget for a given item? Can we detect *sleeping beauties* (Garfield 1980), i.e., items that have the potential to become popular but have yet to?

This work is concerned with three open questions about online promotion and popularity. The first gap is modeling attention at the individual versus the aggregate level. Most previous work study diffusion processes at the individual level, in applications like profiling (Bleier and Eisenbeiss 2015), personalized recommendation (Zhang et al. 2014) or optimizing reach (Zarezade et al. 2017). However, this question remains open: **How to quantify the aggregate attention received by a given item?** Especially, how can we estimate the amount of attention induced by external promotions. The second question is about predicting popularity under promotion. There exists a number of generative models for the popularity of online items (Shen et al. 2014) or of diffusion cascades (Zhao et al. 2015; Mishra, Rizoiu, and Xie 2016), however these concentrate mainly on the prediction accuracy, and they leave open a question important for designing promotion campaigns: **What factors should popularity forecast take into account?** The third question concerns building cost-effective promotion schedules. Solutions have been proposed for building promotion schedules at the individual user level (Spasojevic et al. 2015; Karimi et al. 2016) – i.e. the *when-to-post* problem, when should a user post the content in order to maximize the audience within her social network. However, these approaches do not generalize to an advertisement context, where additional factors should be accounted for, such as repeated promotions, and the monetary value of promotions. The open question is: **What promotion schedules are effective for different content?**

In this work, we answer all three questions above, building upon a recently proposed popularity model, the Hawkes Intensity Process (HIP) (Rizoiu et al. 2017). HIP models popularity as being continuously driven by exogenous stimuli, which may or may not be amplified through an endogenous word-to-mouth diffusion.

Addressing the first question, HIP models attention at the collective-level, by taking the expectation over user behavior in the online. We use it to quantify the expected attention series generated by external promotion, and we propose in Sec. 3 two metrics: the *viral potential* $\nu$ quantifies the total expected views generated from one unit of promotion; the *maturity time* $t^*$ quantifies the time needed for the majority (e.g. $95\%$) of these views to unfold. These serve to determine the amount of promotion needed to achieve a target number

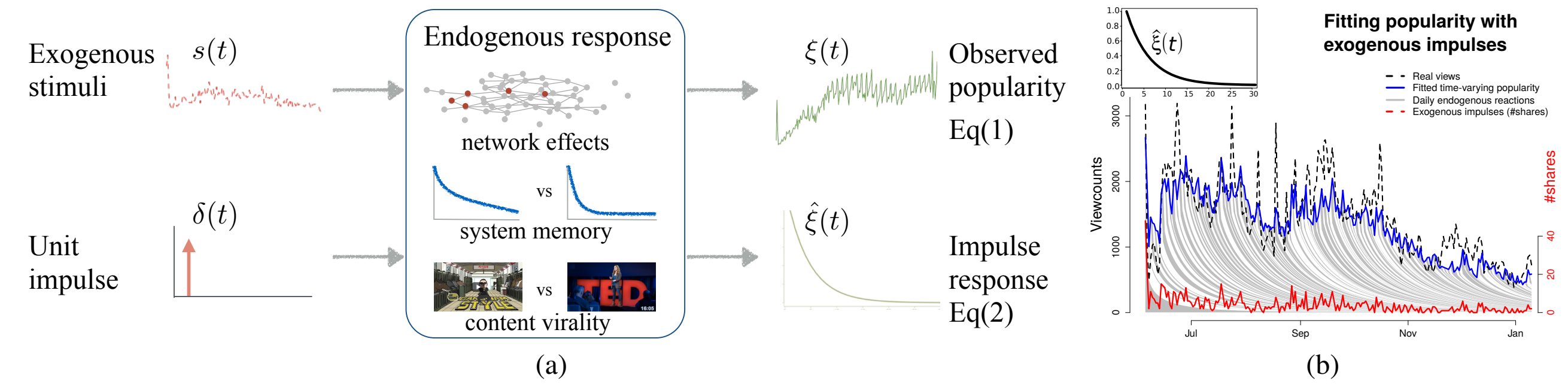

Figure 1: (a) Conceptual schema of the Hawkes Intensity Process. Top flow: popularity series $\xi(t)$ is a result of external stimuli $s(t)$ driving endogenous response involving network effects, system memory and content virality. Bottom flow: the impulse response $\hat{\xi}(t)$ is a result of a stimuli of unit impulse driving the same endogenous response. (b) An illustration of the LTI property. The decomposition of the fitted viewcount series into scaled and temporally shifted *impulse responses* (inset), for the Music video id `2V6_1VxiBt8`.

of views, and comparatively analyze videos.

Addressing the second question, we investigate four factors that influence the forecast of future popularity under promotion: the item's viral potential, the amount of and the timing of promotions, and the non-linear relationship between attention and popularity. We perform in Sec. 5 two predictive exercises. We first forecast future views, and we find that accounting for future promotion volume leads to the largest reduction in prediction error (by 38.66%), whereas the knowledge of promotions timing seems negligible. In the second exercise, we detect items with a sudden rise in popularity. We find that jointly taking into account the viral potential and the non-linearity of popularity performs best.

To answer the third question, we set out to analyze the interplay of maturity time and promotion schedule. We consider promotions to have a time-sensitive value, under the common setting of compounding interests. Our analysis shows that the optimal schedule depends on the items maturity time: the most cost-effective way to promote fast unfolding content is to introduce promotions late; the trade-off between maturity and cost becomes increasingly important for content with slow unfolding.

**The main contributions of this work include:**

- Quantify popularity under promotion with two metrics: the total attention generated by one unit of promotion and the time needed for it to unfold;
- Analyze the factors influencing popularity forecasts, and evaluate prediction of popularity jumps;
- Quantify economic value versus time in promotion, and provide guidelines for constructing cost-effective promotion schedules;
- Explain why constant promotion is desirable – that this schedule increases the perceived viral potential.

**Differences between the current work and HIP.** This work differs from HIP in several significant ways. Firstly, we address the problem of estimating future popularity in the context of planned promotion. In contrast, HIP analyses popularity using naturally occurring external attention, which lacks control in the quantity and timing of intended promotions. Secondly, in this paper one key focus is detecting items which exhibit a sudden jump in popularity. Whereas the original HIP work are mainly concerned with forecasting the amount of future views. Lastly, this work address two new problems – quantifying the factors that affect the prediction of popularity, and profiling economically efficient promotion schedules. The R code for estimating and simulating HIP, and the ACTIVE dataset are publicly available at `https://github.com/andrei-rizoiu/hip-popularity`.

## 2 Prerequisites: The HIP model

In this section, we briefly review the Hawkes Intensity Process (HIP) that was recently introduced (Rizoiu et al. 2017) to model the evolution of popularity under the continuous influence of external stimuli (Sec. 2.1) and we elaborate on the key property of HIP of being a Linear Transform Invariant (LTI) system (Sec. 2.2).

### 2.1 The Hawkes Intensity Process

Self-exciting point processes are a class of stochastic processes, where the occurrence of past events makes the occurrence of future events more probable. They capture several key intuitions of online information dissemination, and recently became one of the defacto choices for modeling social media (Zhao et al. 2015). One such process, the Hawkes point process (Hawkes 1971), uses the following event rate $\lambda(t)$ to measure how likely a viewing event will occur at time $t$:

$$\lambda(t) = \mu s(t) + \sum_{t_i < t} \phi_{m_i}(t - t_i) \quad (1)$$

$\lambda(t)$ has two additive components. The first component is proportional to a measure of external influence $s(t)$ scaled by a constant $\mu$. Here $s(t)$ represents the volume of promotion over time. $\phi_{m_i}(t - t_i)$ is a time-decaying triggering kernel (which here decays following a power-law), representing the rate of views triggered by a previous event $i$, having occurred at time $t_i$ with magnitude $m_i > 0$. Furthermore, each event $t_i < t$ adds to $\lambda(t)$ independently.

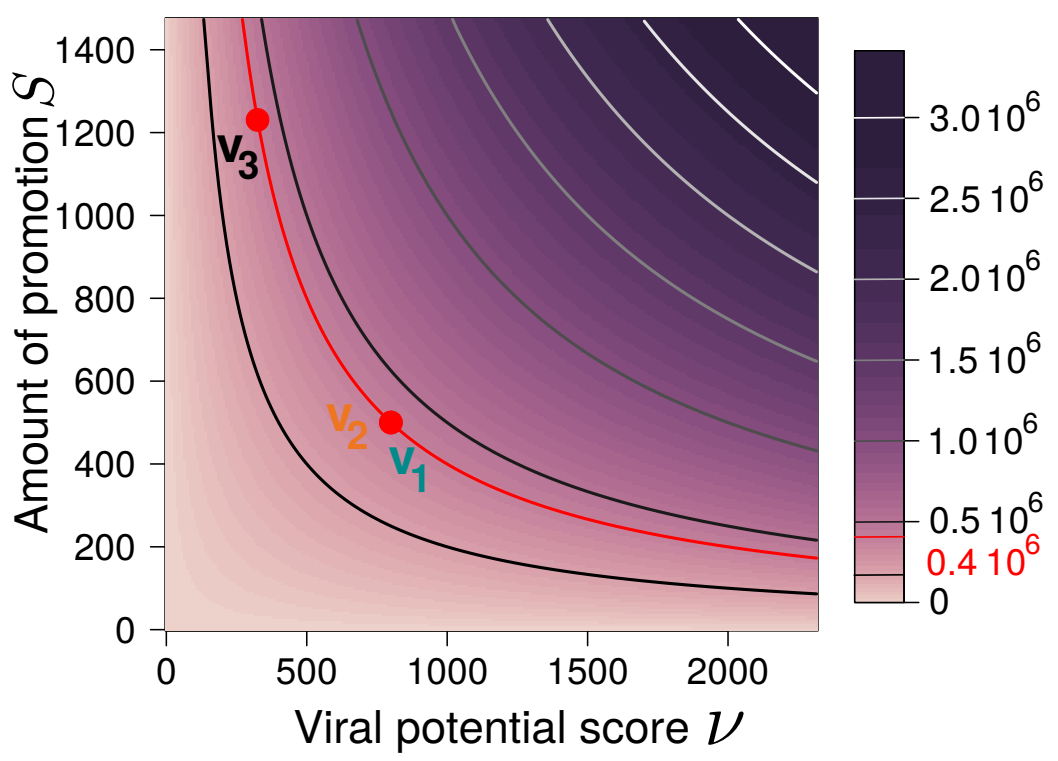

(a)

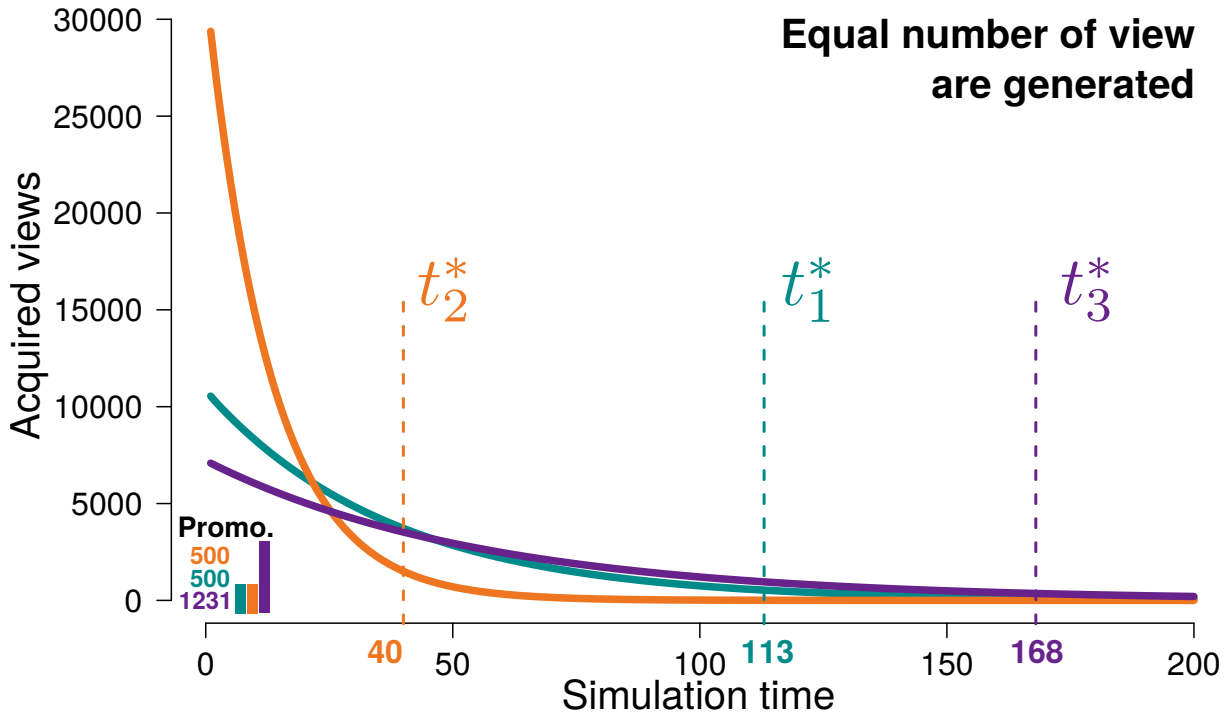

(b)

Figure 2: The relation between viral potential, volume of promotions and gained attention. (a) 3 videos placed on the equal-attention hyperbola $\nu S = 400,000$: $v_1$ (an Entertainment video, id `5TQAnGjyN9A`), $v_2$ (Pets & Animals, id `00ATf2HR-FA`) and $v_3$ (Music, id `bx8O62Tuzno`). (b) 400,000 views are generated by 500 promotions (for $v_1$ and $v_2$) and by 1231 promotions (for $v_3$), placed at $t = 0$. The corresponding maturity time is shown for each video.

HIP (Rizoiu et al. 2017) is derived from marked Hawkes self-exciting processes (Hawkes 1971). It presents a new analytical form that relates the expected number of events (over all possible event history $\mathcal{H}_t$) to the volume of promotions, expressed as the following self-consistent integral equation:

$$\xi(t) := E_{\mathcal{H}_t}[\lambda(t)] = \mu s(t) + C \int_0^t \xi(t-\tau)(\tau + c)^{-(1+\theta)} d\tau \tag{2}$$

Here $\xi(t)$ is the intensity, or the number of expected events in unit time; $\theta$ is a power-law exponent defining the decay of social memory – the larger $\theta$ is, the faster an video is forgotten. $C$ scales the endogenous reaction by taking into account content quality and network properties, and $c$ is a cutoff term to keep the endogenous reaction bounded as $\tau \downarrow 0$. The event intensity $\xi(t)$ is determined by the external stimulus $s(t)$, as well as the event intensity at a previous time $\xi(t-\tau)$ scaled by a corresponding memory kernel $(\tau + c)^{-(1+\theta)}$ for all temporal offsets $\tau < t$. The parameters $\{\mu, \theta, C, c\}$ are estimated for each video from its popularity history (Rizoiu et al. 2017).

## 2.2 HIP as an LTI system

HIP describes a linear time-invariant (LTI) system, as shown previously (Rizoiu et al. 2017). That is to say that if $\xi(t)$ is the event intensity function for input $s(t)$, then the event intensity function for a shifted and scaled version of the input $as(t - t_0)$ is $a\xi(t - t_0)$ for $a > 0, t_0 \geq 0$, i.e., scaled and shifted by the same amount. Consequently, the endogenous reaction for a given video is completely characterized by the event series spawned from one unit of input, denoted as Dirac delta function $\delta(t)$. This response $\hat{\xi}(t)$ is called the *impulse response* of the linear system:

$$\hat{\xi}(t) = \mu\delta(t) + C \int_0^t \hat{\xi}(t-\tau)(\tau + c)^{-(1+\theta)} d\tau \ . \tag{3}$$

Fig. 1a illustrates this process of endogenous-generated popularity driven by exogenous signals – input $s(t)$ results in popularity series $\xi(t)$ (top), whereas an impulse input $\delta(t)$ result in response $\hat{\xi}(t)$. The *sliced* graph in Fig. 1b illustrates an important application of the LTI property. Namely, the popularity series is explained as superpositions of scaled impulse responses (gray and white slices) over time.

# 3 Quantifying popularity under promotions

In this section, we propose two new measures to quantify a video's total *viral potential* and the time it takes to (almost) reach such potential. We reveal one novel insight about the effect of promotions – that timing does not matter in infinite time. Finally, we show how to more accurately estimate attention resulted from promotions, in finite time.

**Viral potential** $\nu$ is designed to capture the total amount of attention generated by one unit of promotion. It is defined as the integral of the impulse response $\hat{\xi}(\tau)$ over infinite time, namely:

$$\nu = \int_0^\infty \hat{\xi}(\tau) d\tau \ . \tag{4}$$

$\nu$ exists and is finite when the branching factor of the HIP $\frac{C}{\theta c^\theta} < 1$. Throughout this paper, we use estimates of $\nu$ by simulating the impulse response series over 10,000 timesteps and numerically integrating it. $\nu$ can be used to compare how promotable videos are relative to each other. A corollary of HIP's LTI property shows a way of using $\nu$ to estimate the return of promotion campaigns.

**Corollary 3.1.** *Any promotion series $s(t), t \in [0, T]$ so that $\int_0^T s(\tau)d\tau = S$, applied to a HIP model with the same parameter will generate $\nu S$ views, given infinite time.*

That is to say that the total amount of attention driven by a given amount of external promotion is fixed, no matter when the promotions happen. This is easy to see with the sliced graph in Fig. 1b, as moving one unit of promotion from any time $t$ to another time $t + \Delta$ results in moving one unit of impulse response by $\Delta$. This does not change the total amount of resulting attention the when $t \to \infty$.

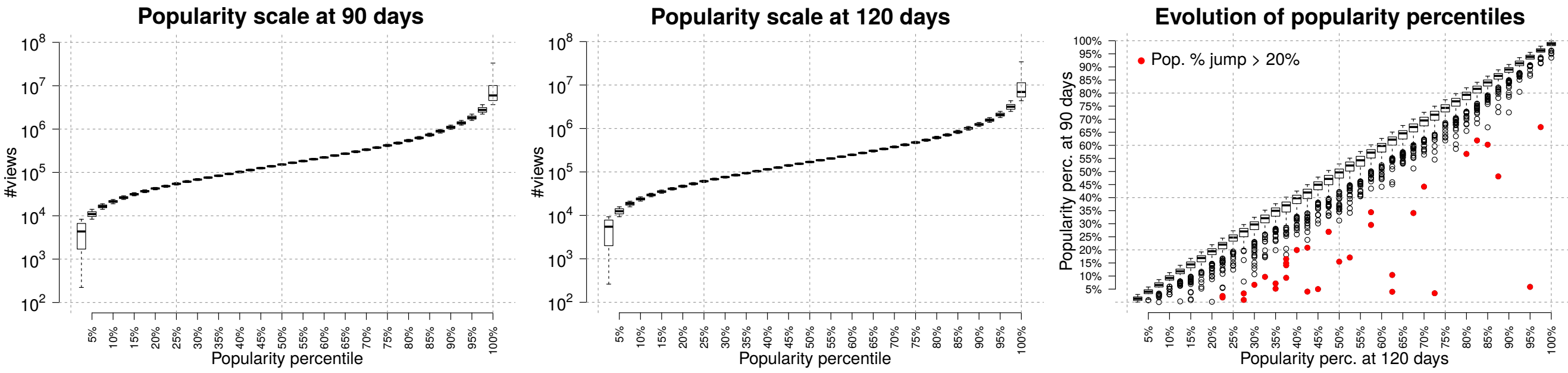

(a) (b) (c)

Figure 3: Popularity scale on the ACTIVE dataset at 90 days (a) and 120 days (b), each box in the box plot contains 2.5%, or 343 videos. (c) Change in popularity between 90 days and 120 days. Outliers are videos whose popularity jumped. The outliers, whose popularity increased by at least 20%, are highlighted in red.

**Maturity time $t^*$.** Having measured the total impact of one unit of promotion, we would also like to quantify how fast a given video achieves such effects. Inspired by the definition of half-life in nuclear physics, we introduce *maturity time* $t^*$, defined as the time it takes to achieve a significant fraction $\omega$ of the total popularity of a unit impulse:

$$t^* = min\left\{ t \geq 0 \middle| \int_0^t \hat{\xi}(t)dt \geq \omega\nu \right\} \quad . \tag{5}$$

$t^*$ is finite when $\nu$ exists and it captures the speed of the decay of $\hat{\xi}(t)$. In this work, we use $\omega = 0.95$, and numerically estimate $t^*$ by performing a linear search in time: we sum the increasingly longer prefixes of $\hat{\xi}(t)$ and we stop when the obtained sum is greater than $\omega\nu$. The timestep where we stop serves as $t^*$.

**The virality and timing of example videos.** We illustrate in Fig. 2 the relation between viral potential, promotion and gained view as a result of promotion. From Corollary 3.1 follows that the hyperbolas $\nu S = const$ define equal-attention lines – i.e. all combinations of potential and promotion volume that amount to the same effect –, which can be used to study videos comparatively. Fig. 2a illustrates this with three example videos, which lay on the same hyperbola corresponding to 400,000 target views. Videos $v_1$ and $v_2$ have the same viral potential of $\nu \sim 800$. They would each require 500 units of promotions to achieve the target. $v_3$ on the other hand, has $\nu \sim 325$ (mainly due to its low exogenous sensitivity $\mu_3 = 5.76$, compared to $\mu_1 = 21.09$ and $\mu_2 = 58.76$), and would require 1,231 units of promotion to attain the same goal. Note that despite having a similar $\nu$, videos $v_1$ has a much longer maturity time $t_1^* = 113$ days than $v_2$, with $t_2^* = 40$ days. This is illustrated in Fig. 2b, which shows generation of the $400,000$ views over time, for the three videos.

**Estimating attention in finite time.** Many applications are time-critical, and it is desirable to estimate the expected attention in finite time. Given a promotion schedule $s(t), t > 0$, and the beginning and the ending of the time period $t_b$ and $t_e$, respectively, we derive from Eq. (3) the expected attention in the time interval $[t_b, t_e]$ as:

$$\Xi[t_b, t_e] = \Xi^{old}[t_b, t_e] + \Xi^{new}[t_b, t_e] \quad \text{, where}$$
$$\Xi^{old}[t_b, t_e] = \int_0^{t_b} s(t) \int_{t_b}^{t_e} \hat{\xi}(\tau - t)d\tau dt$$
$$\Xi^{new}[t_b, t_e] = \int_{t_b}^{t_e} s(t) \int_0^{t_e - t} \hat{\xi}(\tau)d\tau dt \quad . \tag{6}$$

$\Xi^{old}[t_b, t_e]$ corresponds to the views generated by the unfolding of the *old promotions* that occurred before the studied interval. $\Xi^{new}[t_b, t_e]$ corresponds to the incomplete realization of the *current* promotions that occur during $[t_b, t_e]$.

## 4 Popularity of tweeted videos

In this section, we introduce the tweeted video dataset, on which the measurements and the simulations in the rest of this paper are based. We also introduce the popularity scale, and provide a unique observation of videos that *jump*, or suddenly gain popularity after their initial appearance.

**The tweeted videos** dataset was collected (Rizoiu et al. 2017) in 2014, from all tweets that link to a YouTube video. It contains 16 million unique videos tweeted between May and December 2014 and with publicly available popularity stats. In this work we use the ACTIVE subset of the dataset of twitted videos, constructed so that all views, shares and tweets series are at least 120 days long and the video metadata is available. ACTIVE contains 13,738 videos, which were uploaded between 2014-05-29 and 2014-08-09, which were tweeted at least 100 times between 2014-05-29 to 2014-12-26, and were viewed at least 100 times and shared 100 times (through the Youtube share mechanism). Each video has recorded its metadata (title, author, upload date, category) and the daily views, shares and tweets series.

**Popularity scale** is one way to visualize the non-linear relationship between a video's popularity rank (denoted in percentiles) and the number of views it receives (Rizoiu et al. 2017). This scale is estimated from a video collection by sorting videos by their number of views, at a given age. For compactness we group all videos into 40 equal-sized bins, each containing 343 videos in the ACTIVE set. Fig. 3a and 3b plot the boxplots of views in each bin, at 90 and 120 days of age, respectively. Fig. 3c explores the change of popularity percentile between day 90 and 120, denoted as $\Psi_{90}(\cdot)$ and $\Psi_{120}(\cdot)$, respectively. $\Psi_T(v) : \mathbb{R} \rightarrow [0, 1]$ is a function that returns the popularity percentile corresponding to $v$ views, on the popularity scale constructed at age $T$ (i.e. $T$ days after video upload). Most videos stay in or near the same percentile bin (i.e. the high density around the diagonal). The outliers are videos whose popularity "jumped" between day 90 and 120. There are no outliers in the upper-left part of the graphic, as the number of views cannot decrease over time. A video can fall into a lower popularity bin if it accumulates views more slowly than the other videos in his original bin. However, the popularity percentile can increase considerably – Fig. 3c highlights the 33 videos whose popularity jumped by more than 20%. Note that one effect of the non-linear mapping from views to rank is that similar amounts of newly acquired views can have very different impacts on rank: $40,000$ additional views between day 90 and 120 may increase a video's popularity by $15\%$ when starting from lower $17.5\%$ percentile, increase by $2.5\%$ when starting from middle $50\%$ percentile or even decrease by $2.5\%$ when starting from upper $90\%$ percentile – this is because in order to continue to rank in the top $10\%$ at 120 days, a video requires an additional $144,740$ views, more than the $40,000$ being considered here.

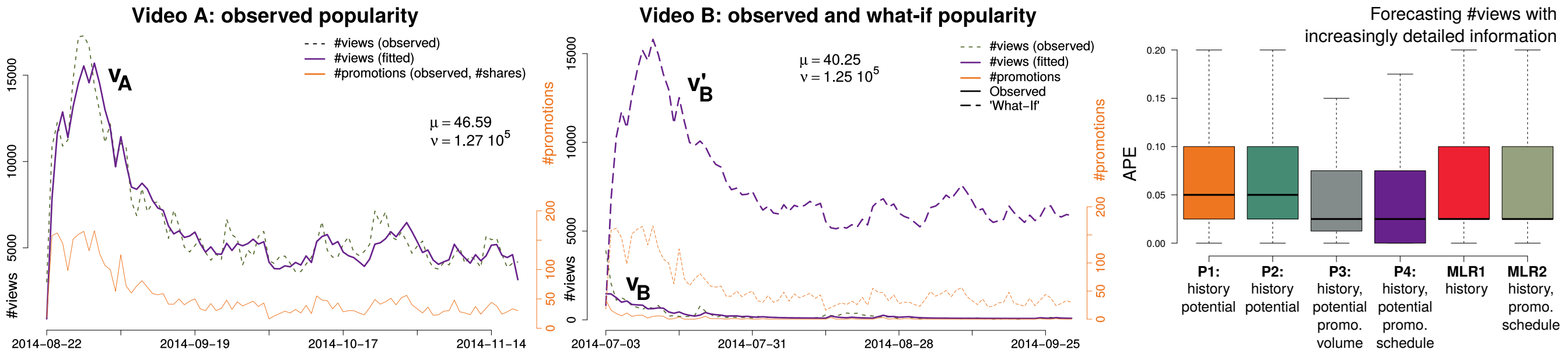

(a) (b)

Figure 4: **(a)** The "What-If" analysis for two videos ($v_A$ and $v_B$) with a similar viral potential, having achieved different popularity levels. (left) The observed and fitted popularity series, the observed promotion and fitted parameters for $v_A$. (right) Observed, fitted popularity series for $v_B$, and the hypothetical series (denoted as $v'_B$), if $v_B$ had had the same promotion as $v_A$. **(b)** APE when predicting the number of views between 91 and 120 days, multiple predictors, using shares as promotions.

## 5 Future popularity under promotion

In this section, we explore the relationship between viral potential, promotion and popularity. Sec. 5.1 illustrates the relationship between promotion and popularity using two example videos that have similar viral potential. HIP is the state of the art in popularity forecasting (Rizoiu et al. 2017), however the factors that influence the performance of the prediction were not well understood. We quantify the relative influence of four factors - the video's viral potential, the amount of promotion, the timing of the promotion and the video's prior position on the popularity scale - in two predictive exercises: predicting future views (Sec. 5.2) and predicting significant rises in popularity (Sec. 5.3).

### 5.1 What-if scenarios for two videos

The relationships among viral potential, promotion, and popularity can be understood in a scenario where we can ask "what-if" certain promotions are applied to a given video, and compare the actual and the hypothetical outcomes. To this end, we select two videos from the ACTIVE dataset that have the same viral potential. Video $v_A$ (id `8jPir15Ms8c`) is an Entertainment video, having achieved a high popularity percentile $\Psi_{120}(v_A) = 85\%$ and it is among the top $2.5\%$ most shared videos, receiving 6192 shares in 120 days. Video $v_B$ (id `WTOC9Lw-gdc`), is a Music video, with a low popularity percentile and few shares ($\Psi_{120}(v_B) = 15\%$, 162 shares). Although the two videos have similar viral potential scores ($\nu \simeq 126,000$), their popularity series present different dynamics (shown in Fig. 4a). The only factor that led to the popularity difference in video $v_A$ and video $v_B$ seems to be the amount of promotions they receive. One may ask "what if" $v_B$ actually went through the same promotions as $v_A$. We simulate the popularity series of video $v_B$ driven by the promotion series of video $v_A$, using Eq. (2). As shown in Fig. 4a right, this "what-if" series is denoted as $v'_B$, and has much more views that $v_B$. The total popularity of $v'_B$ is now almost the same as the popularity of $v_A$. This example shows that given sufficient promotion, an unpopular video with high potential could become popular. This insight is used to predict popularity in the rest of this section.

### 5.2 Predicting future views

In this section, we study the factors which impact the prediction of the number of views received by a video, as a result of promotions. Based on HIP, we construct four predictors, which embed increasingly detailed information about promotions, and we show that this leads to progressively better estimates. More precisely, we study the impact of the amount of promotion and the timing of the promotion. We also study the case when information about future promotions is not available, quantifying how much can be inferred about future popularity based on past popularity only. We denote this latter setup as *prediction*, in contract with *forecasting* the popularity obtained through promotions.

| Predictor | pred. / forecast | Historic Views | Historic Promo. | Viral potential | Future promo. Volume | Future promo. Time | Equation | mean APE ± *std* |
|---|---|---|---|---|---|---|---|---|
| **B** | prediction | ✓ | | | | | $\Psi_{120}^{-1}(\Psi_{90}(V_{1:90}))$ | 16.17%± *14.07%* |
| $\mathbf{P}_1$ | prediction | ✓ | ✓ | ✓ | | | $V_{1:90} + \Xi^{old}[91, 120]$ | 8.37%± *11.17%* |
| $\mathbf{P}_2$ | prediction | ✓ | ✓ | ✓ | | | $V_{1:90} + \Xi^{old}[91, 120] + \nu S_{61:90}$ | 8.12%± *12.07%* |
| $\mathbf{P}_3$ | forecast | ✓ | ✓ | ✓ | ✓ | | $V_{1:90} + \Xi^{old}[91, 120] + \nu S_{91:120}$ | 4.98%± *6.43%* |
| $\mathbf{P}_4$ | forecast | ✓ | ✓ | ✓ | ✓ | ✓ | $V_{1:90} + \Xi^{old}[91, 120] + \Xi^{new}[91, 120]$ | 4.94%± *6.38%* |
| $\mathbf{MLR}_1$ | prediction | ✓ | | | | | | 7.07%± *9.32%* |
| $\mathbf{MLR}_2$ | forecast | ✓ | ✓ | | ✓ | ✓ | | 6.94%± *9.09%* |

Table 1: Summary of predictors used for predicting future views. The columns show the factors included in each predictor, the predicted value (equation) and the obtained prediction performance.

**Experimental setup.** We construct a temporal holdout setup, each video is observed for the first 90 days, on which we fit the parameters of the HIP model, using either tweets or shares as external stimulation. The next 30 days (from 91 to 120) serve as the test period, and the tweets and the shares in this period are used as promotion. We opted for this setup due to the practical difficulty of performing interventions in large social environments, such as Youtube and Twitter. Using tweets and shared during test period as promotions is also supported by the lack of Granger causality between the views, shares and tweets series, shown in (Rizoiu et al. 2017). Commonly-used performance error metrics (such as the root-mean-square-error) are skewed by the long-tailed viewcount distribution shown in Fig. 3, and we chose not to use them. Instead, we evaluate each predictor using the Absolute Percentile Error (APE):

$$APE = |\Psi_{120}(\Xi) - \Psi_{120}(V_{1:120})| \quad , \qquad (7)$$

where $\Xi$ and $V_{1:120}$ are the predicted and the observed number of views during the first 120 days, respectively.

**Constructed predictors.** Table 1 summarizes the four constructed predictors and the three used baselines, showing the information leveraged by each predictor, the predicted value formula (where applicable) and the prediction performance. The predictors $\mathbf{P}_1$ and $\mathbf{P}_2$ operate in a predictive setup, they do not make use of the promotion information. $\mathbf{P}_1$ leverages history (views and attention that the videos received during the training period) as well as the video viral potential. It quantifies the views in the period 91:120 obtained by the unfolding of the "old" promotions (promotion observed during the first 90 days) using $\Xi^{old}[1 : 90]$ (Eq. 6). The predictor $\mathbf{P}_2$ is similar to $\mathbf{P}_1$ and it makes a more realistic assumption about future promotions: it assumes that videos receive during the period 91-120 days the same amount of promotion as in the period 61-90 days (denoted as $S_{61:90}$). The third predictor, $\mathbf{P}_3$, leverages history, viral potential and future promotion volume; it works in a forecasting setting in which the number of promotions in test period are known ($S_{91:120}$). Predictor $\mathbf{P}_4$ leverages history, viral potential and promotion scheduling; in addition to $\mathbf{P}_3$, it leverages the promotion schedule in the test period $s(t), t \in [91, 120]$ and predicts the number of views using the formula in Eq. (6). We show in Table 1 the data baseline **B**, which outputs, for each video, the number of views required for it to maintain, at 120 days, the same popularity percentile as at 90 days. We also implement two baseline predictors, using multivariate linear regression (MLR), based on the observation that past popularity is indicative of future popularity (Pinto, Almeida, and Gonçalves 2013). $\mathbf{MLR}_1$ leverages the past viewcounts to predict the viewcounts during the period 91-120 days. $\mathbf{MLR}_2$ is enhanced with information about promotions, both historic and future. Note that popularity prediction systems that require observing individual events (Zhao et al. 2015; Mishra, Rizoiu, and Xie 2016) cannot be used here, since the views information from Youtube is aggregated daily. $\mathbf{MLR}_1$ and $\mathbf{MLR}_2$ are evaluated with 5-fold cross validation. Shares are used to train the HIP model and $\mathbf{MLR}_2$ (and they serve as promotions in the forecasting setup). The results for tweets are shown in the online supplement (sup 2016).

**Results.** Fig. 4b and Table 1 summarize the performances of our predictors and the baselines, over the ACTIVE dataset. The data baseline **B** shows the difficulty of the problem, all other predictors and baselines outperform this data baseline. Visibly, introducing more detailed information consistently leads to a reduction in mean APE. Accounting for future promotions has the largest impact, reducing mean APE from 8.12% of $\mathbf{P}_3$ to 4.98% of $\mathbf{P}_2$. Further accounting for promotion schedule has a considerably smaller impact, with $\mathbf{P}_4$ outperforming $\mathbf{P}_3$ by only 0.04% in APE. Consistent with the results in (Rizoiu et al. 2017), both HIP-based forecasting $\mathbf{P}_3$ and $\mathbf{P}_4$ consistently outperform the baseline $\mathbf{MLR}_2$. The situation is different for the predictive exercise, where the baseline $\mathbf{MLR}_1$ outperforms both $\mathbf{P}_1$ and $\mathbf{P}_2$. Approximating the future volume of promotions (in $\mathbf{P}_2$) has only marginal impact.

## 5.3 Predicting popularity jumps

Recall in Fig. 3c there are 33 videos that had large *jumps* in the popularity scale. Predicting such sudden increase in popularity is relevant for content optimization or marketing. This can be done by building upon the methods in Sec. 5.2, and taking into account the non-linear effect of the popularity scale. We study the following factors which influence the prediction of popularity: the viral potential, the promotion

| Predictor | Oracle | Viral potential | Promo. volume | Pop. scale position | Equation | AUC |
|---|---|---|---|---|---|---|
| $\mathbf{R}_1$ | | ✓ | | | $\nu$ | 0.6 |
| $\mathbf{O}_1$ | ✓ | | ✓ | | $\frac{V_{91:120}}{S_{91:120}}$ | 0.62 |
| $\mathbf{R}_2$ | | ✓ | ✓ | | $\nu S_{91:120}$ | 0.79 |
| $\mathbf{O}_2$ | ✓ | | | | $V_{91:120}$ | 0.88 |
| $\mathbf{R}_3$ | | ✓ | ✓ | ✓ | $\Psi_{120}(V_{1:90} + \nu S_{91:120})$ $-\Psi_{90}(V_{1:90})$ | 0.91 |
| $\mathbf{O}_3$ | ✓ | | | ✓ | $\Psi_{120}(V_{1:120}) - \Psi_{90}(V_{1:90})$ | 1 |

Table 2: Summary of predictors used for predicting popularity jumps. The columns show the factors included in each predictor, the predictors output value (equation) and the obtained prediction performance.

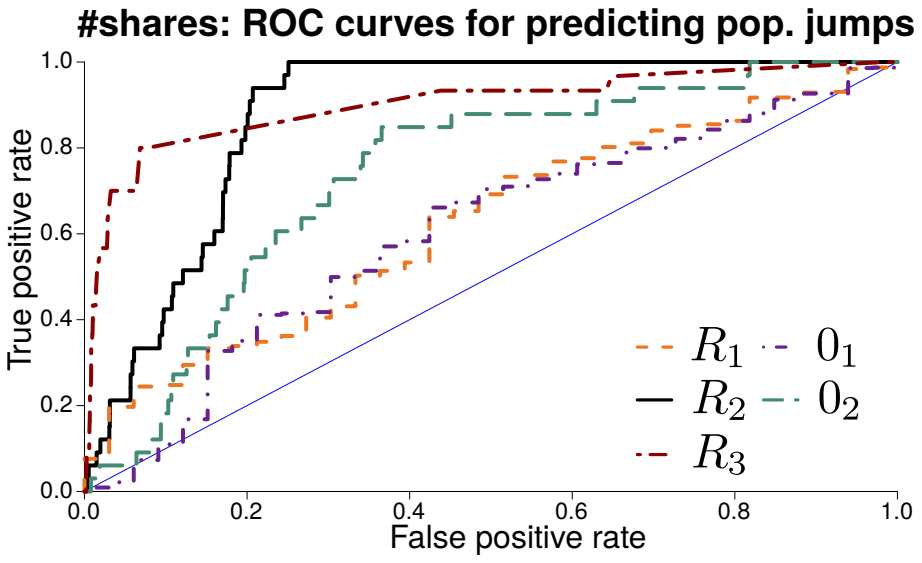

Figure 5: ROC curves of detecting popularity *jumps* – videos with a sudden increase in popularity of more than 20% between day 90 and 120, using shares as promotion.

volume and the position on the popularity scale.

**Constructed predictors.** Based on HIP, we construct three predictors which incorporate increasingly more information. $\mathbf{R}_1$ outputs the video viral potential. $\mathbf{R}_2$ leverages the viral potential and the amount of promotions: it outputs the forecasted number of views during the period 91-120 days. $\mathbf{R}_3$ leverages the viral potential, the amount of promotions and the position on the popularity scale at day 90: it outputs the forecasted popularity percentile gained during the period 91-120 days. Each predictor is doubled by an "oracle" predictor, based on the real data in the test period. The oracle predictors ($\mathbf{O}_1$, $\mathbf{O}_2$ and $\mathbf{O}_3$) serve as baselines to benchmark the maximum performance that can be achieved using a particular information. For example, $\mathbf{O}_1$ mirrors $\mathbf{R}_1$ in using the viral potential alone. Note that the viral potential is the number of views generated by a single promotion, and can be computed on the testing data as $\frac{V_{91:120}}{S_{91:120}}$ (where $V_{91:120}$ and $S_{91:120}$ are the series of view and shares during 91-120 days). Similarly, $\mathbf{O}_2$ outputs the *observed* number of view in the period 91:120 days, and $\mathbf{O}_3$ the popularity percentile gained in the same period.

**Experimental setup.** We employ the same temporal hold-out setup as in Sec. 5.2: each video in the ACTIVE dataset is observed during the first 90 days, and the "jumping" behavior if predicted on the next 30 days. All predictors output real values, which we transform into a binary classifier using a threshold. We vary the threshold and we report the performance using ROC curves and AUC (Area Under Curve). Note that $auc \in [0.5, 1.0]$, higher is better, the random binary classifier obtains $auc = 0.5$ and perfect classification scores $auc = 1$. We varied the popularity jump threshold from 0.1 to 0.5 and the trends and the conclusions are similar for all values. Here, we show the results for a jump threshold of 0.2, the rest are in the online supplement (sup 2016).

**Results.** Fig. 5 shows the ROC curves obtained for each predictor, using shares as promotions. The results for tweets are shown in the online supplement (sup 2016). The AUC obtained by each predictor is shown in Table 2. Both $\mathbf{R}_1$ and the oracle $\mathbf{O}_1$ achieve rather low prediction performances, with $auc(\mathbf{R}_1) = 0.60$ and $auc(\mathbf{O}_1) = 0.62$. This indicates that video potential is only weakly linked to jumping behavior, which happens due to a number of reasons: videos of low potential that receive high promotion, or videos with high potential that receive low promotion, and the varying effects of views across the popularity scale. Accounting for the volume of promotion provides a boost in prediction performance, both for $\mathbf{R}_2$ ($auc(\mathbf{R}_2) = 0.79$) and for the oracle ($auc(\mathbf{O}_2) = 0.88$). Note that the oracle $\mathbf{O}_2$ does not achieve perfect performance in predicting popularity jumps. This is due to the unequal impact of views on popularity, e.g. in order to jump 20% on the popularity scale, a video needs 83,245 views if it started from 25%, and 327,820 views if it starts from the median 50%. We obtain a performance boost when we account for the position on the popularity scale at day 90, with $auc(\mathbf{R}_3) = 0.91$. Note that the forecasts made using potential, promotion and scale information ($\mathbf{R}_3$) are more accurate than the oracle $\mathbf{O}_2$, achieving a 0.66 true positive rate with 0.05 false positive rate. This demonstrates the importance of taking into account the non-linear effect of the popularity scale. $\mathbf{O}_3$ obtains a perfect score ($auc(\mathbf{O}_3) = 1$).

## 6 Profiling promotion schedules

In this section, we focus on time as a key variable for deploying promotions and observing their outcome, and present two novel observations. The first one explores the trade-off between maturity time of a video and the time-sensitive value of promotions under the economic setting of compounding interest (Sec 6.1). In the second, we explain the common practice of advertising at regular intervals as an increase in perceived potential (Sec. 6.2).

### 6.1 The economics of time for promotions

Following from Corollary 3.1, the scheduling of $S$ promotions does not matter in infinite time, as they eventually results in $\nu S$ units of popularity gain. However, practical applications seek to optimize the impact of promotions after a given time, e.g., how much was a video's popularity boosted, a month after the promotion campaign has ended? The two counter-acting factors at play here are both about time – the time that the effects of a promotion have to unfold –, and the time-sensitive (monetary) value of carrying out promotions for a given piece of content. If the cost of promotion at any give time is constant, then the best strategy is to place all promotions as early as possible, this will lead to maximum unfolding of popularity in the target time window. A more realistic setting, however, needs to account for the varying value of promotions in carried out in different times. One simple way to model this is via compounding interest (Witt 1613) with rate $a$; say every dollar stored for one unit time (e.g. day) is worth $(1 + a)$ dollars. In this setting, putting all promotions at the beginning of $k$ days actually needs $(1+a)^k$ times the cost of putting all promotions at the last day.

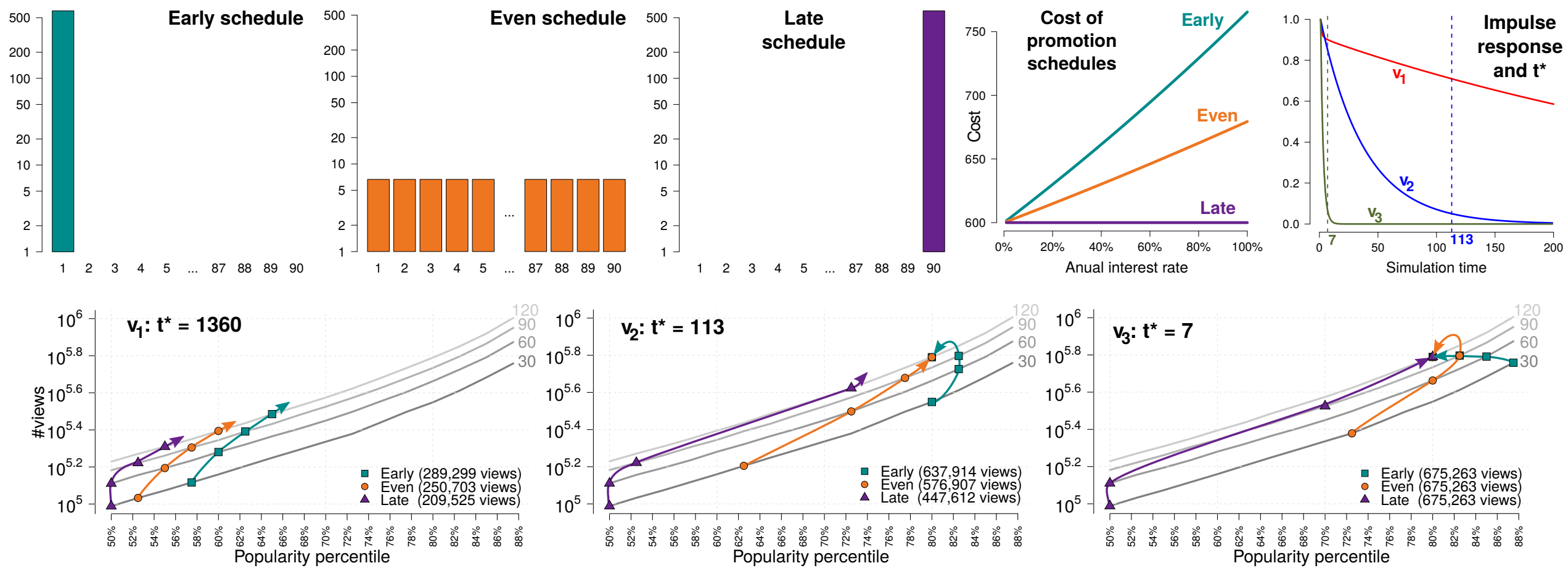

Figure 6: Analysis of three promotion schedules on the achieved popularity, after 120 days. (top) Distribution of 600 promotions over 90 days, for the **early**, **even** and **late** schedules (the y-axis is log-scale). The cost of schedules, function of interest rate. The impulse response and maturity time for the three subject videos: $v_1$ (id `9MwgExiIjnY`) is a video of a Japanese youtuber, $v_2$ ( id `5TQAnGjyN9A`) is a short Entertainment video, $v_3$ (id `d03B1Zp4GtQ`) is a Brazilian boy gang video about playing games and drinking. (bottom) Reaction of each of the three videos to the promotion schedules.

**The cost of three prototypical schedules.** We first discuss three prototypical promotion schedules – late, even and early. Without loss of generality, we set the promotion budget at 600 promotions and the promotion time window at 90 days. **late** is spending the promotion budget at the last day of the promotion period. This is the least costly option, i.e. 600\$. **even** is spreading the promotion budget equally each day. Its cost is $\frac{600}{90}\sum_{k=1}^{90}(1 + a)^k$, visibly higher than the cost of late. **early** is spending the entire promotion budget in the first day, for a cost of $600(1 + a)^k$. These three schedules are general enough to provide an intuition of the trade-off between maturity and value of currency. An arbitrary schedule can be composed from them and its effects are also linear combinations of what is presented here, due to the LTI property (Sec. 2.2). The top row of Fig. 6 illustrates each of the three prototypical schedules and the cost of each schedule versus the annually interest rate.

**Scheduling effects on three videos.** We simulate the effects of each of the three promotion schedules on three videos and we plot the results for each video in Fig. 6 (bottom). The median values of the popularity scale at 30, 60, 90 and 120 days (as seen in Fig. 3) are shown as increasingly lighter gray lines. We plot the position (i.e. the popularity bin) obtained by the videos at each of these temporal checkpoints, when promoted with the three schedules. Here, we set each simulated video at median (50%) popularity and mandate that it remain at the median at 30, 60, 90, and 120 days if there were no promotions. The three videos have similar viral potential ($\nu \simeq 800$), but with very different maturity time $t^*$ (shown in Fig. 6 (top right) ): $t^* = 1360$ days for $v_1$, $t^* = 113$ days for $v_2$ and $t^* = 7$ days for $v_3$. For video $v_1$, the slow unfolding of promotion effect makes the early schedule the most effective as the other schedules never catch-up. For video $v_2$, the even schedule amounts to the same popularity percentile at 120 days as the early schedule. Not all promotions in the even schedule completely unfold, and the return of the two schedules is not exactly the same, however the 80% popularity bin covers volumes of view between 574,007 and 762,478 views, and both videos get assigned to it. Notice also that the early promotion achieves a popularity of 82% at 60 and 90 days, but is degraded to 80% at 120 days. This is because, after $t^*$ days, promotions stop generating new views, and the video's popularity slips. This phenomenon is more obvious for $v_3$, for which the even schedule catches-up with early at 90 days, and all three schedules achieve the same popularity percentile at 120 days, but for different costs.

We conclude that promotion strategies should vary for videos of different maturity time: late schedule is the most economical for videos whose popularity unfolds fast enough; for videos with a long maturity time, early promotion is costly but will lead to the most attention.

## 6.2 Promotion and perceived potential

In this section, we study in more detail the **even** promotion schedule. Despite not being the most advantageous in terms of cost or popularity unfolding, even promotion schedules are worth further examination, since they are very widely used in commercial settings – the car advertisement before the prime-time TV show every evening, for example. The process is shown in Fig. 7: we show in graph (a) the example of a common popularity pattern of Youtube videos: the daily viewcount has an initial peak, followed by a steady decay.

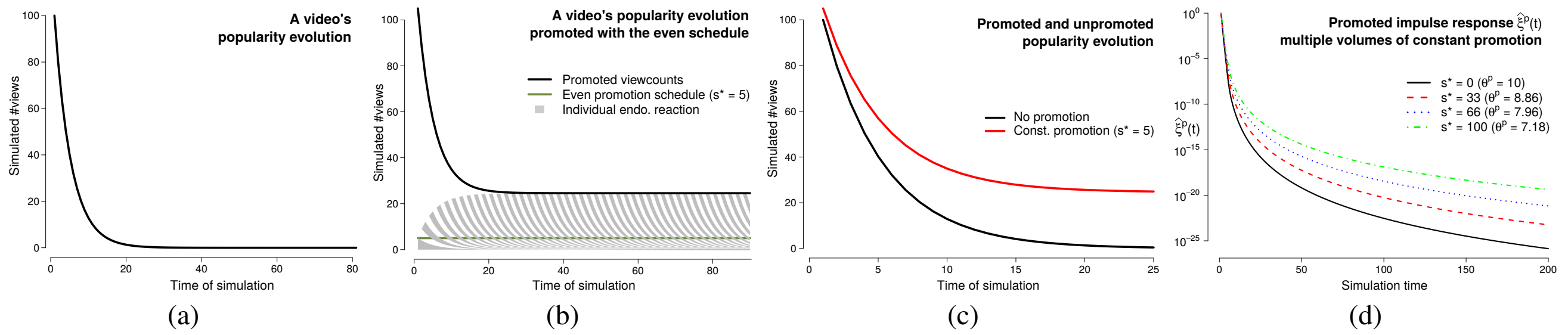

Figure 7: Effects of the even promotion schedule: apparent increase of video viral potential. (a) An example viewcount evolution; (b) the same evolution under an even promotion schedule with $s^* = 5$ promotions daily. (c) Side-by-side plotting of the two evolutions. (d) Equivalent *promoted* input response $\hat{\xi}^p(t)$, for increasing amounts of constant promotion.

Graph (b) shows the same system constantly promoted with a daily promotion of $s(t) = s^* = 5, \forall t \in 1, 2, \dots 90$. With $t^* = 14$, the number of additional views gained through promotion stabilizes after about 20 days. The number of additional daily views is $\nu s^*$ due to the LTI property (Sec. 2.2), as shown with the sliced graph (Rizoiu et al. 2017). Fig. 7 (c) contrasts the promoted and the unpromoted relaxation; we see that the daily views in the promoted system have a slower decrease and converge to a non-zero value – the evenly promoted system appears to posses a longer memory. We refit the HIP model parameters of the promoted system with varying $s^*$ (without informing HIP about the promotion), as shown in Fig. 7 (d). We observe that the equivalent memory exponent $\theta^p$ is smaller and the *promoted* impulse response $\hat{\xi}^p(t)$ has a slower decay. Both are indicative of a longer system memory, and a higher viral potential. This provides a plausible explanation for the widely used even promotion schedule: by constantly injecting promotion into a social system, the perceived viral potential appears increased and the video stays longer in the public attention, thus generating more views.

## 7 Related work

This work is related to three topics in this area: measurement of online social networks, models for popularity and especially stochastic point processes, and the effects of external promotion.

**Measurements of one or more social networks.** There are a number of well-known measurement studies of content, user, and popularity on social networks, including Twitter (Kwak et al. 2010), YouTube videos (Gill et al. 2007), news media (Leskovec, Backstrom, and Kleinberg 2009), as well as specific measurements about popularity and auxiliary attributes such as locality (Brodersen, Scellato, and Wattenhofer 2012). Most current measurement studies are on a single social media network, or done independently for different networks; our work observes behavior from two different networks (Youtube and Twitter) linked by the same media item. A few studies link two networks for predictive tasks, including the novel user-level measurements done by Abisheva *et al.* (Abisheva et al. 2014), a study by Yu *et al.* (Yu, Xie, and Sanner 2014) on using Twitter feeds to predict Youtube popularity change and a system by Yan *et al.* (Yan, Sang, and Xu 2014) for finding optimal Twitter followees to maximize video promotion on Twitter. Our popularity scale over time that are inspired by these large-scale measurements, we then use this scale to measure the performance of popularity models.

**Modeling attention and popularity.** A number of models have been proposed to describe the volume of social media activity over time. The seminal meme-tracker (Leskovec, Backstrom, and Kleinberg 2009) uses a curve with polynomial increase followed by exponential decay to describe sawtooth-shaped volume of news mentions. SpikeM (Matsubara et al. 2012) uses a fixed memory component, modulated by a periodic component, however it does not explicitly account for external influence. Most recently, Tsytsarau *et al.* (Tsytsarau, Palpanas, and Castellanos 2014) models popularity volume as the convolutions two sequences, news event importance and media response, each of a predefined shape. (Raghavan et al. 2014) models online activity patterns using a couple Hidden Markov Model. Stochastic point-processes provide a formal language to describe social events over time, and is seen in a number of state-of-the-art models. These include: reinforced Poisson processes (Shen et al. 2014), multi-scale survival processes (Zhang et al. 2016), self-excited processes used to predict popularity of retweet cascades (Zhao et al. 2015; Mishra, Rizoiu, and Xie 2016). The version of Hawkes processes averaged over an entire network and event times is the HIP model (Rizoiu et al. 2017) that we build upon. This work focus on predicting popularity jumps and quantifying the effects of promotion schedules based on point processes.

**Scheduling promotions.** Intervention strategies for individual and group attention has been a long-standing topic, going back to as long as marketing existed. (Sun 2005) found that endogenous consumption responds to promotion as a result of forward-looking and stockpiling behavior. In the online, promotion studies concentrated on user profiling and personalized online advertising (Bleier and Eisenbeiss 2015) or personalized product recommendation (Zhang et al. 2014). (Chierichetti, Kleinberg, and Panconesi 2014) determines the sequence of activating nodes so as to maximize the total activation in a network. (Liu, Slotine, and Barabási 2011) apply linear control theory to observe that sparse inhomogeneous networks are difficult to control (or steer to a desired state). In broadcast advertising, (Bollapragada, Bussieck, and Mallik 2004) use a mixed-integer program to schedule multiple airings of a television advertisement as evenly as possible. In this work, we provide a mathematically-grounded method to quantify the effects of promotion schedules, and provide key intuitions why evenly-spaced advertising may be preferred.

## 8 Discussion

We present a mathematically-grounded model to estimate the amount of attention driven by external promotion. We adopt a stochastic point process model for popularity, which allows us to propose two novel metrics relating to the promotability of Youtube videos: the viral potential, designed to capture the total amount of attention generated by one unit of promotion and the maturity time, measuring the speed with which this attention is accumulated. We show that we can accurately predict dormant videos – i.e. videos that exhibit a sudden increase in popularity, at a later time after the initial appearance. We perform a cost-return analysis of the impact of promotion schedule on popularity and we explain the widely used constant promotion strategy as increasing perceived social memory.

**Assumptions, limitations and future work.** This work makes a number of simplifying assumptions, some of which can be address in future work. First, we assume that the system's reaction to external random stimuli is identical to the reaction to promotion. We leave as future work to measure how users respond to a post if it were from a friend, or from a publicity source on the social web. Second, the HIP model describes the average behavior of the network (Rizoiu et al. 2017), and we assume this does not change over time, e.g. in the 120 days studied here. Future work can include modeling local network topology and its temporal shifts. Lastly, this paper assumes that popularity ranking is the optimization target. Content- and category-specific ranking are also valid targets, and are left as future work.

**Acknowledgments** This material is based on research sponsored by the Air Force Research Laboratory, under agreement number FA2386-15-1-4018. We thank the National Computational Infrastructure (NCI) for providing computational resources, supported by the Australian Government. We thank Alban Grastien and Christian Walder for insightful discussions that seeded ideas in this work.

# **Appendix:** Online Popularity under Promotion: Viral Potential, Forecasting, and the Economics of Time

Marian-Andrei Rizoiu and Lexing Xie

## 1 Easily promotable and unpromotable videos in the ACTIVE set

In the main text Sec. 5.1, we present the analysis of a "what-if" scenario which shows that, given enough promotion, an unpopular video with high viral potential could become popular. We denote such videos as *easily promotable videos*. Conversely, *non-promotable videos* are videos with low provability score, that do not respond well to promotion on the selected channel. The early identification of videos in these two categories has a direct real-live application, by maximizing the yield of online marketing campaigns.

**Profiling viral potential per category.** In Fig. 9, we break down per category the distribution of the obtained viral potential score $\nu$. The top row corresponds to the results obtained using `#shares` as exogenous impulse, and the bottom row corresponds to the results using `#tweets`. We see that the categories `Shows`, `Gaming` and `Howto&Style` are the most promotable when using either `#shares` or `#tweets`. Similarly, `Nonprofit&Activism` and `News&Politics` seem the least promotable through either channels. Both results seem intuitive, as the relevance of `News&Politics` decays rapidly (i.e. they are unpromotable by nature), while the `Gaming`-related ecosystem has seen a consistent increase in received attention over the recent years.

**Unpromotable videos.** We define as unpromotable a video with $\nu < 1$, in other words, a video for which each promotion generates less than one view. 3798 videos in ACTIVE ($\sim 27.07\%$) are unpromotable through `#shares` channel and 4577 videos ($\sim 32.62\%$) are unpromotable through `#tweets`. Fig. 9 (middle column) shows the percentage of non-promotable videos for each category. Consistent with previous results, `Nonprofit & Activism` and `News & Politics` observe the highest percentages on non-promotable videos. We speculate that this is related to the ephemeral character and short-term relevance of these videos.

**Easily promotable videos** are videos with high viral potential score $\nu$, that have achieved only modest popularity popularity because they have not received enough attention. This is equivalent to discovering future celebrities: individuals with high potential, still to achieve success. One method to identify easily promotable videos is to look for videos which simultaneously obtain a high viral potential score (in percentiles, here above 60%) and a low popularity percentile (here below 40%). 315 videos ($\sim 2.24\%$) are in this category when considering `#shares` as external influence and 334 ($\sim 2.38\%$) using `#tweets`. Fig. 9 (right column) show the percentage of easily promotable videos in each category, for `#shares` (top) and `#tweets` (bottom). The two promotion environments appear less coherent with respect to easily promotable videos, by comparison to unpromotable videos. For example 26 `Film & Animation` videos are easily promotable through `#shares`, but only 13 videos in this category are unpromotable using `#tweets`.

## 2 Effect of even promotion schedule on real videos.

In the main text Sec. 6.2, we provide an explanation for the effectiveness of the even promotion schedule: the perceived increase of the viral potential of the promoted video. In this section, we further study the impact of the even promotion schedule on real videos from the ACTIVE dataset.

Figure 10(a) shows a colormap representing the viral potential score as a function of $\mu$ and $A_{\hat{\xi}}$. Note that the axis of this colormap are different from those of the map presented in the main text Fig. 2a. Three `Music` videos are depicted on the map: video $v_1$ (id `6Dhk08_cVjg`) belongs to a Nigerian singer, and it exhibits a high exogenous sensitivity $\mu$, but a low endogenous response $A_{\hat{\xi}}$; video $v_2$ (id `7A8Wlkyr99c`) is of an artist from Martinique, which has a low exogenous sensitivity and a high endogenous response; video $v_3$ (id `3ZJunhnBylY`) is from a Mexican artist and has both high $\mu$ and high $A_{\hat{\xi}}$. The viral potential scores of both $v_1$ and $v_2$ are consistently lower than that of $v_3$. The other three graphics in Figure 10 detail the popularity evolution and each system's reaction to even promotion. All three videos show similar popularity evolutions, with an initial burst of views, followed by a decreasing evolution. For each video, we denote the observed (in black) and the simulated (in red) popularity series, as well as the promoted (in blue) popularity series with a constant promotion of $s^* = 1$. The gray bars show the additional views gained as a result of promotion, and their height converges to $\nu s^* = \nu$ (here $s^* = 1$).

# 3 Additional plots

In this section, we show the additional plots referenced in Sec. 5.2 and 5.3 in the main text: a) Fig 8 shows the results of predicting future views with increasingly detailed information, when training the HIP model using `#tweets` as source of promotion; b) Fig 11 shows the ROC curves of the performance of predicting "jumping behavior", when the popularity jump threshold is varied between $0.1$ and $0.5$. Table 3 shows the AUC (Area under ROC curve) obtained by the same predictors.

Table 3: The performance (measured using AUC) when predicting videos with a popularity jump. The threshold of what is considered a "popularity jump" is varied between $0.1$ and $0.5$ (shown by the table column headers). The predictions are made using either `#shares` or `#tweets` as promotions, shown in the top and bottom part of the table, respectively. The construction of the predictions $\mathbf{R}_1$, $\mathbf{R}_2$ and $\mathbf{R}_3$, and of the oracles $\mathbf{O}_1$, $\mathbf{O}_2$ and $\mathbf{O}_3$ is detailed in the main text Sec. 5.3.

| | **pred.** | **0.10** | **0.15** | **0.20** | **0.25** | **0.30** | **0.35** | **0.40** | **0.45** | **0.50** |
|---|---|---|---|---|---|---|---|---|---|---|
| #shares | $\mathbf{R}_1$ | 0.54 | 0.56 | 0.60 | 0.63 | 0.60 | 0.63 | 0.58 | 0.62 | 0.62 |
| #shares | $\mathbf{O}_1$ | 0.54 | 0.58 | 0.62 | 0.57 | 0.55 | 0.60 | 0.59 | 0.53 | 0.53 |
| #shares | $\mathbf{R}_2$ | 0.78 | 0.75 | 0.79 | 0.71 | 0.76 | 0.75 | 0.72 | 0.67 | 0.67 |
| #shares | $\mathbf{O}_2$ | 0.83 | 0.86 | 0.88 | 0.91 | 0.93 | 0.94 | 0.95 | 0.96 | 0.96 |
| #shares | $\mathbf{R}_3$ | 0.90 | 0.91 | 0.91 | 0.86 | 0.93 | 0.91 | 0.92 | 0.89 | 0.89 |
| #shares | $\mathbf{O}_3$ | 1.00 | 1.00 | 1.00 | 1.00 | 1.00 | 1.00 | 1.00 | 1.00 | 1.00 |
| #tweets | $\mathbf{R}_1$ | 0.60 | 0.52 | 0.54 | 0.60 | 0.69 | 0.75 | 0.79 | 0.81 | 0.81 |
| #tweets | $\mathbf{O}_1$ | 0.57 | 0.57 | 0.56 | 0.64 | 0.68 | 0.64 | 0.54 | 0.64 | 0.64 |
| #tweets | $\mathbf{R}_2$ | 0.77 | 0.71 | 0.71 | 0.68 | 0.64 | 0.60 | 0.58 | 0.52 | 0.52 |
| #tweets | $\mathbf{O}_2$ | 0.83 | 0.86 | 0.88 | 0.91 | 0.93 | 0.94 | 0.95 | 0.96 | 0.96 |
| #tweets | $\mathbf{R}_3$ | 0.82 | 0.79 | 0.84 | 0.79 | 0.82 | 0.80 | 0.81 | 0.76 | 0.76 |
| #tweets | $\mathbf{O}_3$ | 1.00 | 1.00 | 1.00 | 1.00 | 1.00 | 1.00 | 1.00 | 1.00 | 1.00 |

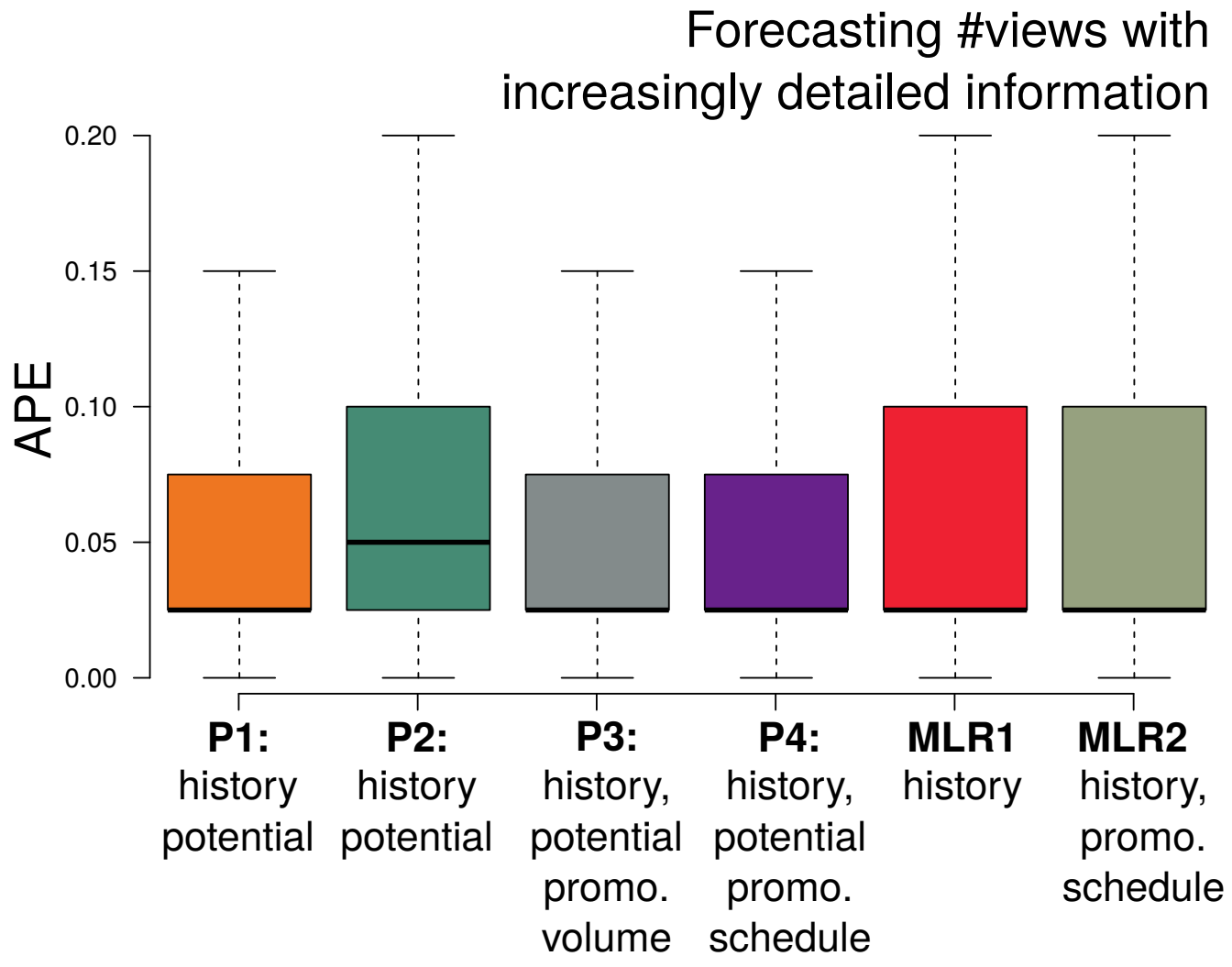

Figure 8: APE when predicting popularity acquired between 91 and 120 days, using tweets as the promotion series. The description of the predictors $\mathbf{P}_1$, $\mathbf{P}_2$, $\mathbf{P}_3$, $\mathbf{P}_4$ and the two baselines $\mathbf{MLR}_1$ and $\mathbf{MLR}_2$ is given in Sec. 5.2 of the main text.

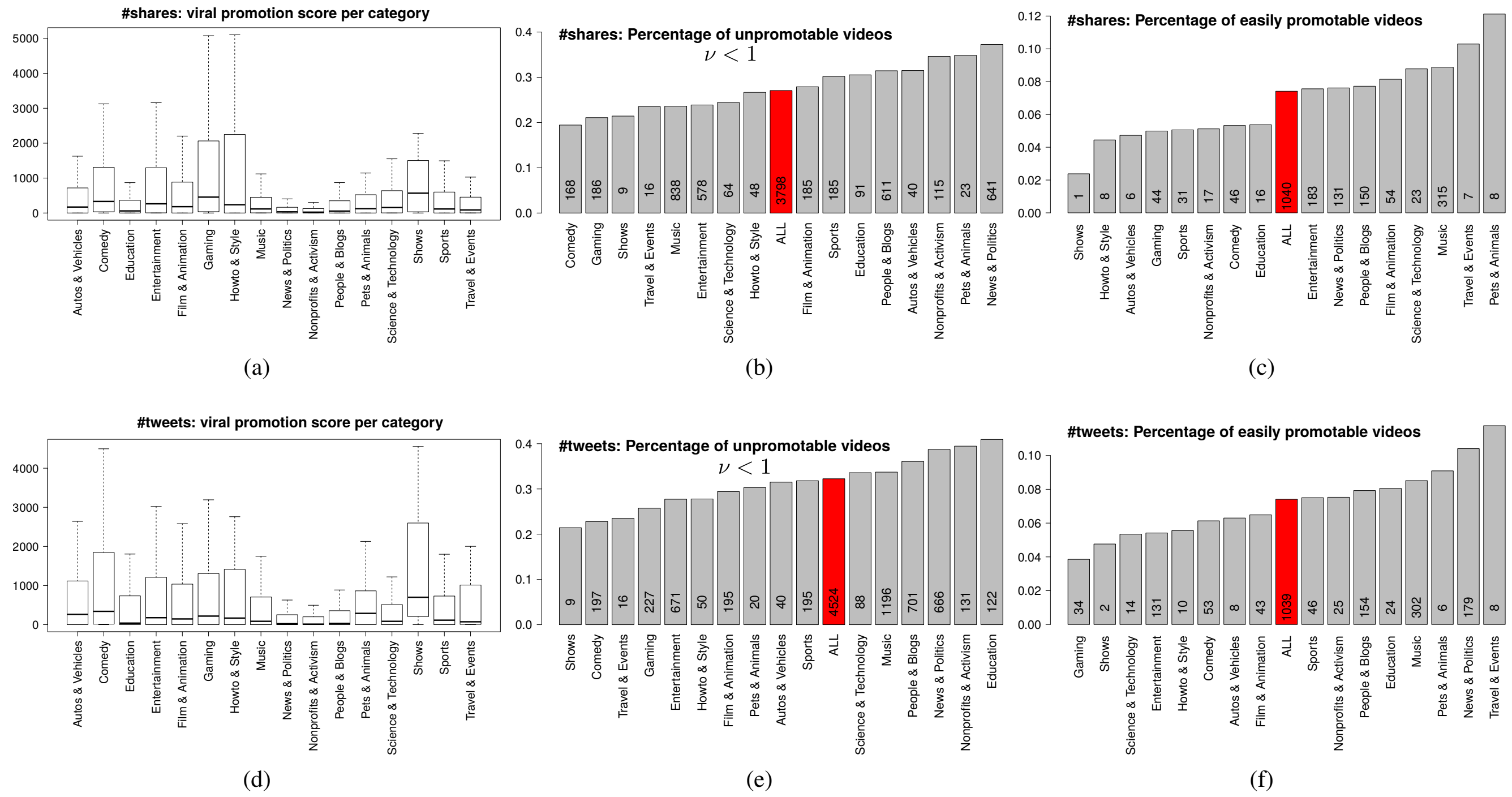

Figure 9: (left column) Breakdown of promotion score over categories, shown as boxplots. (middle column) Percentage of unpromotable videos in each category ($\nu < 1$). The bars are ordered ascendantly by their height. The bars corresponding to categories are shown in gray color and the one for the entire collection (denoted by ALL) is shown in red color. The value on each bar indicates the absolute effectives for each category. (right column) Percentage of easily promotable videos in each category. The top row shows the values obtained when trained using `#shares` as external influence, while the bottom row corresponds to `#tweets`.

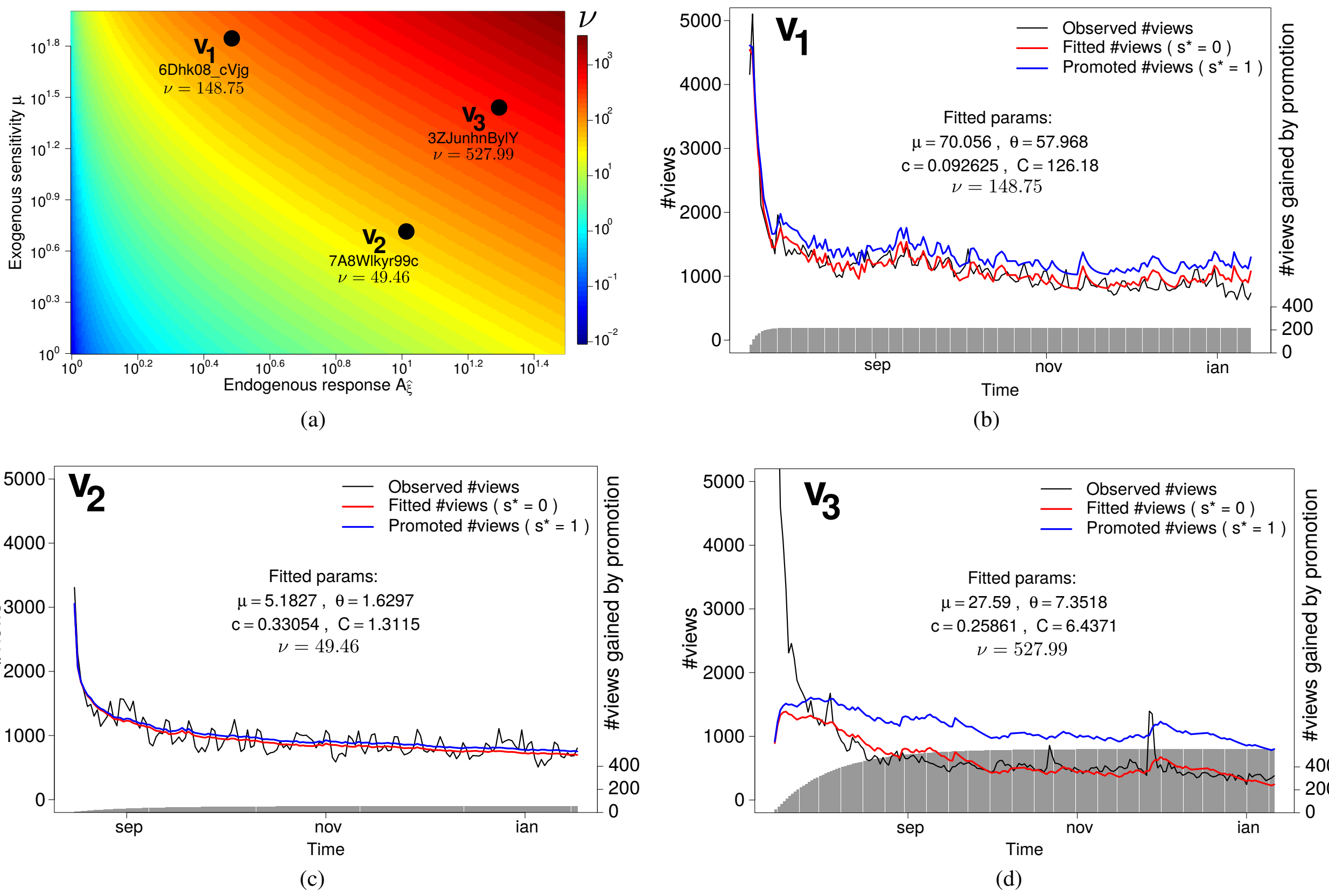

Figure 10: (a) Color map visually depicting the viral potential score as a function of the endogenous response and of the exogenous sensitivity. Three real-life videos ($v_1$ id `6Dhk08_cVjg`, $v_2$ id `7A8Wlkyr99c`, and $v_3$ id `3ZJunhnBylY`) are depicted on the map. (b-d) The observed time-varying popularity is presented using black lines, and it shows for all three videos a similar evolution: an initial popularity followed by a decay. The red line depicts the fitted viewcount series and the blue line shows the promoted viewcounts series, under the even promotion schedule using $s^* = 1$ one unit of promotion. The bars at the bottom of the graph show the amount of gained views by introducing the constant exogenous influence.

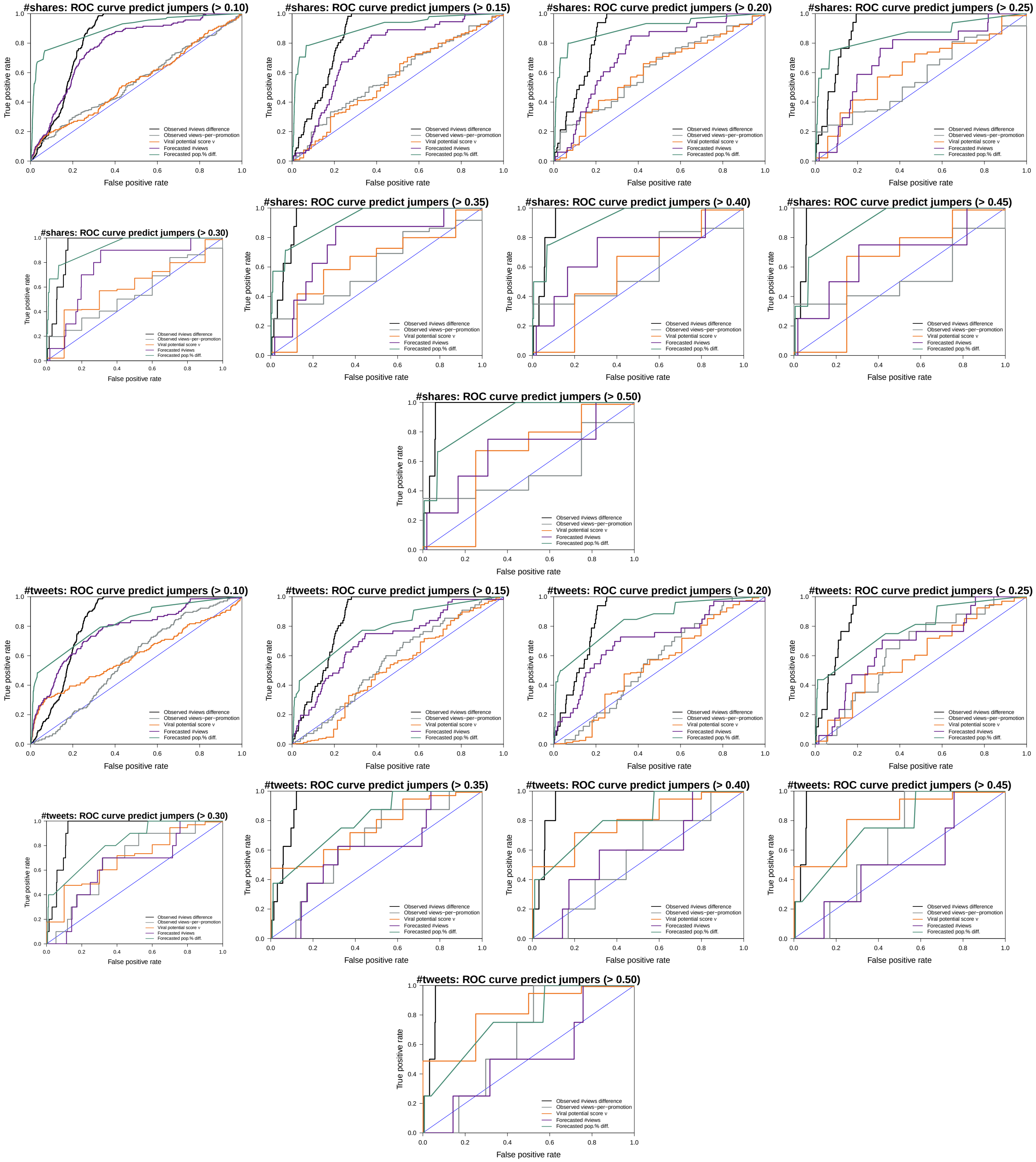

Figure 11: ROC curves for predicting the whether a video's popularity will "jump" for `#shares` (top 3 rows) and for `#tweets` (bottom 3 rows). The popularity jump threshold is varied between 0.1 and 0.5, with a step of 0.05.